\documentclass[adp, fleqn]{w-art}
\usepackage{makeidx}
\usepackage{graphicx}
\usepackage{dcolumn}
\usepackage{bm}
\usepackage{amsmath}
\usepackage{amssymb}
\def \be{\begin{equation}}
\def \ee{\end{equation}}

\input{tcilatex}

\begin{document}

\title{Josephson junctions as detectors for non-Gaussian noise}

\author[B. Huard]{B. Huard\inst{1}}
\address[\inst{1}]{Quantronics group, Service de
Physique de l'\'Etat Condens\'{e}, DRECAM, CEA-Saclay, 91191
Gif-sur-Yvette, France}
\author[H. Pothier]{H. Pothier\inst{1,}\footnote{Corresponding author\quad E-mail:~\textsf{hugues.pothier@cea.fr}}}
\author[Norman O.\ Birge]{Norman O.\ Birge\inst{1,}\footnote{Permanent address: Department of Physics and Astronomy,
Michigan State University, East Lansing, Michigan 48824-2320, USA}}
\author[D.\ Esteve]{D.\ Esteve\inst{1}}
\author[X. Waintal]{X.\ Waintal\inst{2}}
\address[\inst{2}]{Nanoelectronics group, Service de Physique de l'\'Etat Condens\'{e}, DRECAM, CEA-Saclay, 91191
Gif-sur-Yvette, France}
\author[J. Ankerhold]{J.\ Ankerhold\inst{3}}
\address[\inst{3}]{Institut f\"ur Theoretische Physik, Universit\"at Ulm, Albert-Einstein-Allee 11, 89069 Ulm, Germany}

\begin{abstract}
Non-Gaussian fluctuations of the electrical current can be detected with a Josephson junction placed on-chip with the
noise source. We present preliminary measurements with an NIS junction as a noise source, and a Josephson junction in
the thermal escape regime as a noise detector. It is shown that the Josephson junction detects not only the average
noise, which manifests itself as an increased effective temperature, but also the noise asymmetry. A theoretical
description of the thermal escape of a Josephson junction in presence of noise with a non-zero third cumulant is
presented, together with numerical simulations when the noise source is a tunnel junction with Poisson noise.
Comparison between experiment and theory is discussed.
\end{abstract}

\maketitle

\section{Introduction}

The fluctuations of the electrical current reveal the charge of the
carriers, their correlations as well as their fermionic nature.
Until recently, only the variance of the current fluctuations had
been measured and predicted. Theoretical progress in the last decade
allowed one to calculate the full counting statistics of electrons
passing through any phase coherent circuit made of conductors such
as tunnel junctions, quantum point contacts,
diffusive wires or chaotic cavities\cite%
{Levitov,Nazarov,BelzigNazarov,Nagaev,Pilgram}. For short enough measurement times, the distribution of the number of
transmitted charges is predicted to differ sizeably from a Gaussian distribution. On the experimental side, few
experiments have gone further than measuring the average noise, i.e. the quadratic average of the fluctuations. Indeed,
observing a departure from the Gaussian distribution of the fluctuations requires the measurement of small signals in a
short time with a high accuracy. The pioneering experiment performed by Reulet \textit{et al.} \cite{Reulet} used
analog microwave techniques to reconstruct $\left\langle V^{3}\right\rangle ,$ with $V$ the voltage across a
current-biased, $\sim 50\,\Omega $-resistance tunnel junction. This first measurement of the third order cumulant of
the voltage fluctuations (or skewness) pointed out the importance of feedback effects associated with the
electromagnetic environment of a noise source\cite{Been,ReuletHouches}. Bomze \textit{et al.} \cite{Bomze} performed a
direct counting of single tunneling events across a tunnel junction, using a high-precision analog-to-digital
converter. Another approach consists in using an on-chip detector\cite{Hei}. The statistics of the charge in quantum
dots in the classical regime of sequential tunneling was probed with a quantum point contact \cite{Ensslin,Fujisawa}.
It was also shown that the current through a Josephson junction in the Coulomb blockade regime is sensitive to
non-Gaussian noise \cite{Lindell}. A promising scheme to probe current fluctuations, proposed by Tobiska and Nazarov
\cite{Tobiska}, consists in using a Josephson junction as a threshold detector. The switching of a Josephson junction
out of the zero-voltage state provides a sensitive threshold detector because the switching rate varies exponentially
with the bias current. We present here a preliminary experiment in which a current-biased Josephson junction is
capacitively coupled to a voltage-biased tunnel junction which induces current fluctuations in the Josephson junction.
A similar experiment was performed by Timofeev \textit{et al.}~\cite{Pekola expt} using a junction in a regime
involving macroscopic quantum tunneling, thermal escape and phase diffusion. The response of the Josephson junction to
the squewness of current fluctuations was compared with an adiabatic model which uses as an input the measured response
to the average current and as a fit parameter a bandwidth, of the order of the plasma frequency. In our experiment, the
junction is kept in the regime of thermal escape which allows direct comparison to theory. The switching rate of the
Josephson junction is measured for opposite values of the average bias current Ib through it, thus giving access to the
asymmetry of the current fluctuations. In a second part, theoretical predictions for the response of the Josephson
junction to fluctuations with a non-zero third cumulant are presented, under the assumption that the correlation time
of the fluctuations is shorter than the response time of the Josephson junction, and that feedback effects related to
the voltage developing on the Josephson junction can be neglected. The results of this calculation are then compared
with numerical simulations that integrate the Langevin equation describing the phase dynamics in presence of current
fluctuations created by uncorrelated tunnel events. Finally, experimental results are confronted with theory.

\section{Measuring noise with a Josephson junction}

In a simplistic picture, a Josephson junction biased at an average current $%
I_{b}$ below its critical current $I_{0}$ can be seen as a threshold
detector for current fluctuations: neglecting thermal activation or quantum
tunneling, the junction remains in its zero-voltage state as long as the
current fluctuations are smaller than $I_{0}-I_{b}.$ When the current
exceeds this limit, the junction switches to the dissipative state, and a
voltage develops.
\begin{vchfigure}[tbph]
\begin{center}
\includegraphics[width=7.5cm]{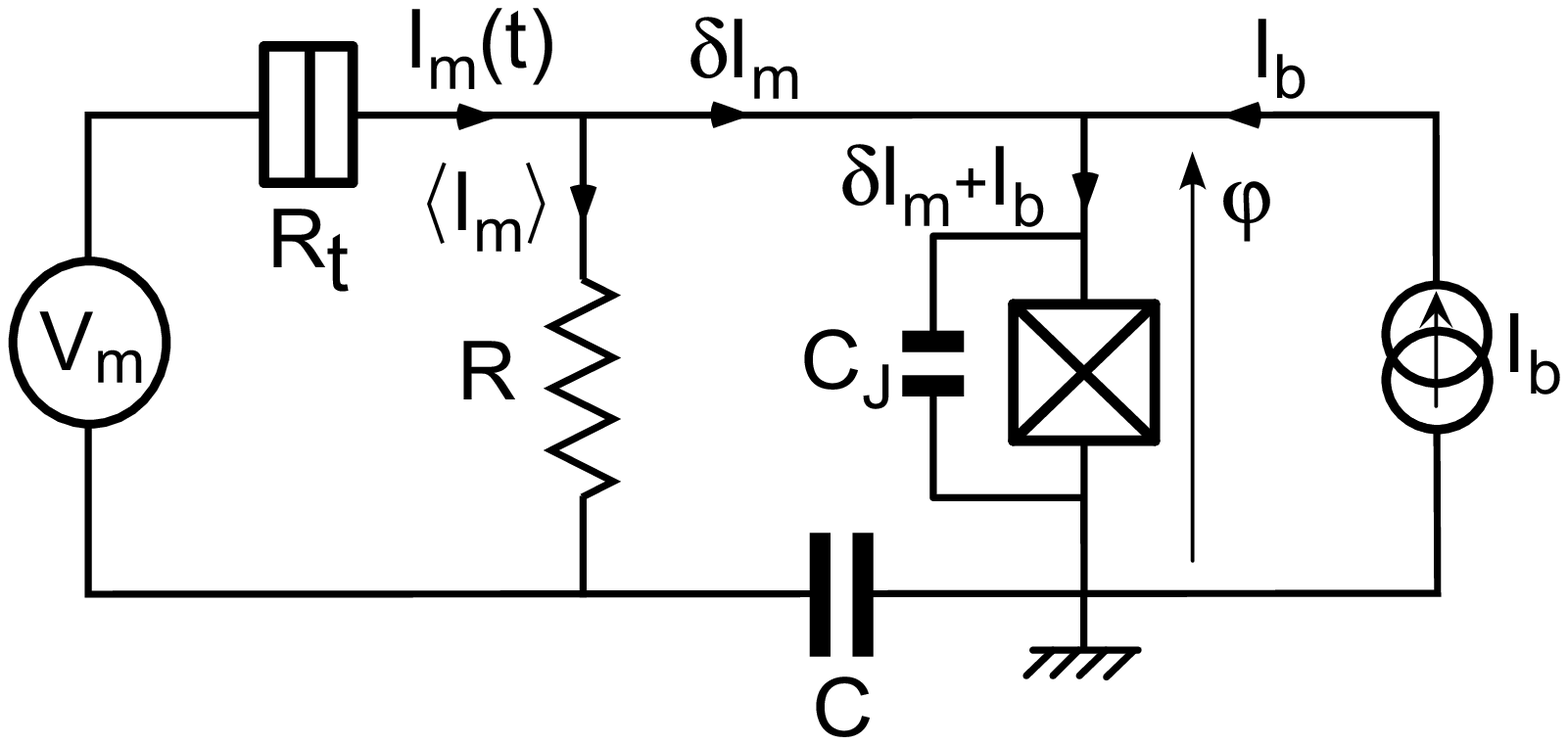}
\end{center}
\vchcaption{Simplified scheme of the experiment. The high frequency ($%
f>1/RC $) fluctuations $\protect\delta I_{m}$ of the current through
a voltage-biased tunnel junction pass through a Josephson junction.
The switching of the Josephson junction to the finite-voltage state
depends on the sum of $\protect\delta I_{m}$ and of the pulsed
current bias $I_{b}(t)$. The switching probability of the junction
during one pulse can be related to the current fluctuations.}
\label{SchemeEPS.eps}
\end{vchfigure}
This voltage subsists until a ``reset'' is performed, by driving the bias
current down to zero, then back to $I_{b}.$ The behavior of the Josephson
junction is thus similar to that of an electrical circuit breaker. The
principle of our experiment is to couple, through a capacitor, the current
fluctuations $\delta I_{m}$ of a voltage-biased noise source to a
current-biased Josephson junction (see Fig.~1), and to compare the switching
rates $\Gamma _{+}$ and $\Gamma _{-}$ at opposite values of the bias current
$\pm I_{b}$ applied during a time $\tau _{p}$ through the Josephson
junction. The asymmetry in the rates $R_{\Gamma }=\Gamma _{+}/\Gamma _{-}-1$
reflects the difference in the probabilities that during $\tau _{p}$
fluctuations such that $\delta I_{m}>I_{0}-I_{b}$ and $\delta
I_{m}<-(I_{0}-I_{b})$ occur, \textit{i.e.} the asymmetry in current
fluctuations whose amplitude exceeds $I_{0}-I_{b}$.

However, a Josephson junction is not a perfect threshold detector:
the switching of a Josephson junction occurs for bias currents
$I_{b}$ smaller than $I_{0},$ because of thermal activation or
quantum tunneling. As explained below, thermal activation is the
dominant process in our experiment; we will thus consider this case
only. Any dissipative impedance connected to the Josephson junction
produces Johnson-Nyquist current noise, which can yield thermal
escape and switching. The theoretical description of this phenomenon
is based on the Resistively Shunted Junction (RSJ) model, in which
the dynamics of the Josephson junction is similar to that of a
particle in a tilted washboard potential with friction
\cite{Barone,Likharev}. The state of the Josephson junction is
characterized by the phase difference $\varphi $ across the
junction, which relates to the position of the fictitious
particle. The particle evolves in the Josephson potential $%
-E_{J}(\mathrm{cos}~\varphi + s\varphi )$, where $E_{J}=I_{0}\varphi
_{0}$ is the Josephson energy, with $\varphi _{0}=\hbar /2e,$ and
$s=I_{b}/I_{0}$ the reduced bias current, which determines the
average slope of the potential. At small bias current, the particle
is trapped in a local minimum and oscillates at the plasma frequency
$\omega _{p}(s)=\omega _{0}(1-s^{2})^{1/4} $, where the frequency
$\omega _{0}\simeq (I_{0}/\varphi _{0}C_{J})^{1/2}$ is set by the
parallel capacitance $C_{J}$. When the current increases further,
the thermal energy can become comparable to the barrier height
$\Delta U(s)\approx (4\sqrt{2}/3)\,\varphi _{0}I_{0}(1-s)^{3/2}$.
The particle can then be thermally activated over the
barrier and escape from the local minimum at a rate%
\begin{equation}
\begin{array}{ccc}
\Gamma (s)\simeq  & \underbrace{\frac{\omega _{p}(s)}{2\pi }} & \underbrace{%
\mathrm{e}^{-\varDelta U(s)/k_{B}T_{\mathrm{esc}}}} \\
& A & \mathrm{e}^{-B}%
\end{array}
\label{rate}
\end{equation}%
In this expression, the escape temperature $T_{\mathrm{esc}}$ is a measure
of the current noise in the circuit connected to the Josephson junction,
mostly at the plasma frequency. In the setup presented in Fig.\thinspace 1,
two noise sources are connected to the Josephson junction: the resistance $R$
in series with the tunnel junction, that can be assumed to be at the bath
temperature $T_{0},$ and the tunnel junction itself, whose current noise is
described by a spectral density $2e\left\langle I_{m}\right\rangle \coth
(eR_{t}\left\langle I_{m}\right\rangle /2k_{B}T_{0}),$ with $R_{t}$ the
tunnel resistance. In a simple model, the escape temperature in
Eq.\thinspace (\ref{rate}) is therefore given by \cite{theseBH}:
\begin{equation}
T_{\mathrm{esc}}=(1/R+1/R_{t})^{-1}(T_{0}/R+e\left\langle
I_{m}\right\rangle /2k_{B}\coth (eR_{t}\left\langle
I_{m}\right\rangle /2k_{B}T_{0})).  \label{teff de noise}
\end{equation}

In this setup, the Josephson junction is thus very sensitive to the second
moment of the current fluctuations\cite{Deblock,Pekola seul}.
\begin{vchfigure}[htbp]
\begin{center}
\includegraphics[width=10cm]{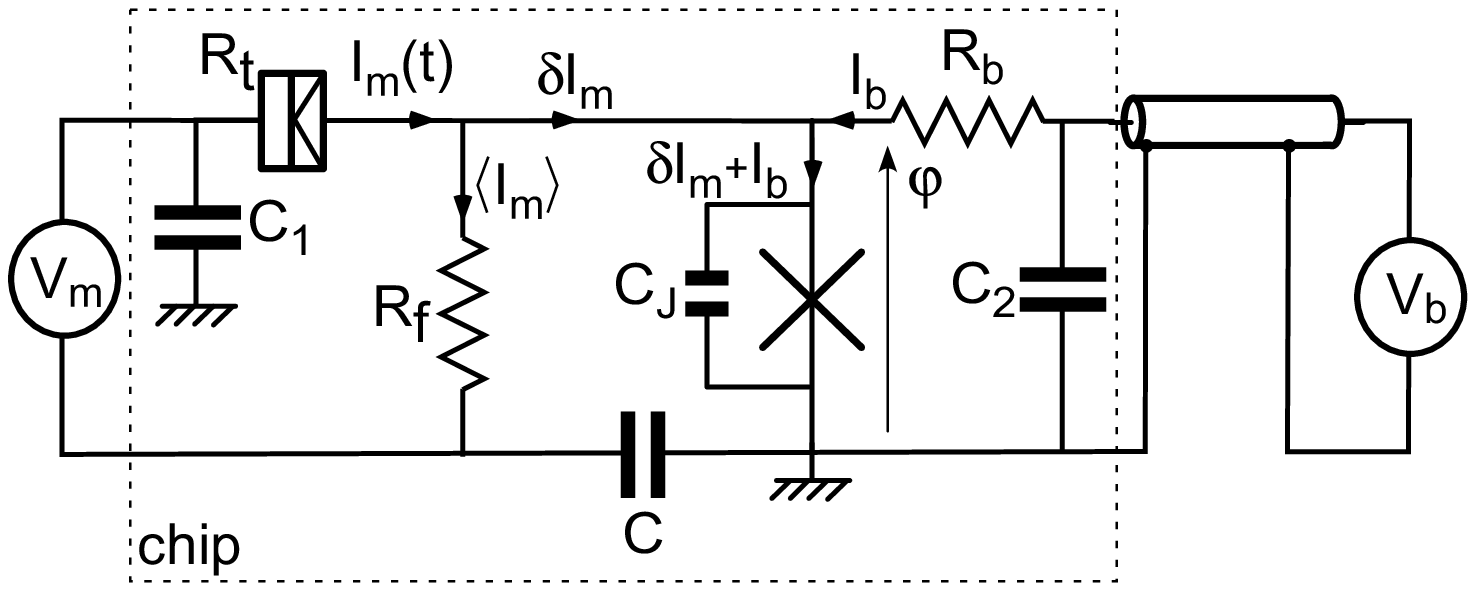}
\end{center}
\vchcaption{Full scheme of the experiment. The noise source is a
Cu-Al NIS tunnel junction. The current pulses through the Josephson
junction are obtained by voltage pulses on a coaxial line, applied
to the series combination of a bias resistor $R_b$ and the Josphson
junction. Large capacitors $C$, $C_1$ and $C_2$ allow for good
voltage bias and limit the impedance seen by the Josephson junction
to on-chip elements. } \label{realSchemeEPS}
\end{vchfigure}

The measurements presented here were obtained on a sample with an aluminum
Josephson junction of area $0.7\,\mu \mathrm{m}^{2}$, in parallel with a
capacitor $C_{J}=36\,\mathrm{pF.}$ For convenience in sample fabrication,
the noise source under investigation was not a normal (NIN) tunnel junction,
but a $0.2\,\mu \mathrm{m}^{2}$ NIS junction made of Al and Cu electrodes
separated by a tunnel barrier obtained simultaneously with the barrier of
the Josephson junction by oxidation of the aluminum layer$.$ In order for
the electromagnetic environment of the Josephson junction to be entirely
determined by on-chip elements, large capacitors ($C_{1}\approx 850\,\mathrm{%
pF}$ and $C\approx C_{2}\approx 130\,\mathrm{pF}$) were fabricated on-chip
between the connecting pads and ground (see Fig. \ref{realSchemeEPS}). These
capacitors and the junction capacitor $C_{J}$ were made using plasma
oxidized aluminum electrodes deposited on the substrate prior to patterning
the circuit \cite{theseBH}. At frequencies of the order of the plasma
frequency, the capacitors act as short circuits in parallel with the circuit
placed outside the chip. Capacitance $C_{1}$ also allows for a good voltage
bias of the junction and reduces the low-frequency cut-off of the high-pass
filter associated with $R_{f}$ and $C$.  Resistors $R_{f}=202~\Omega $ and $%
R_{b}=202~\Omega $ were commercial surface mount NiCr resistors
connected directly on the sample with silver epoxy. The sample
holder was thermally anchored to the mixing chamber of a dilution
refrigerator with a base temperature of $20~\mathrm{mK.}$ All lines
connected to the sample were twisted pairs equipped with microwave
filters, except a $50\,\Omega $ coaxial line with distributed
$50\,\mathrm{dB}$ attenuation designed to apply current pulses to
the Josephson junction. The floating bias of the NIS\ tunnel
junction was realized with a battery-operated voltage divider,
followed by an electrical two-position switch that allowed reversals
of the sign of the bias (at $\sim 1\,\mathrm{Hz}$) to eliminate the
effect of drifts. Bias resistors of $10\,\mathrm{M\Omega }$ were
placed on the two wires of the twisted pair connecting the voltage
source to the samples. The voltage was measured across the series
combination of the tunnel junction (noise source) and of the
Josephson junction (detector), allowing for measurements of one or
the other, depending on the bias scheme. For example, the $I-V$
characteristics of the NIS junction shown in Fig.\thinspace
\ref{IVJJ} was measured with $I_{b}=0$, so that the voltage drop
across the
Josephson junction was zero. From this characteristic taken at $20\,\mathrm{%
mK}$, we deduced the tunnel resistance $R_{t}=1.57\,\mathrm{k}\Omega .\;$In
the following, we describe measurements with the NIS\ junction biased above
the gap voltage, in a transport regime dominated by tunneling of single
electrons, like in a usual NIN\ tunnel junction.
\begin{vchfigure}[tbph]
\begin{center}
\includegraphics[width=12cm]{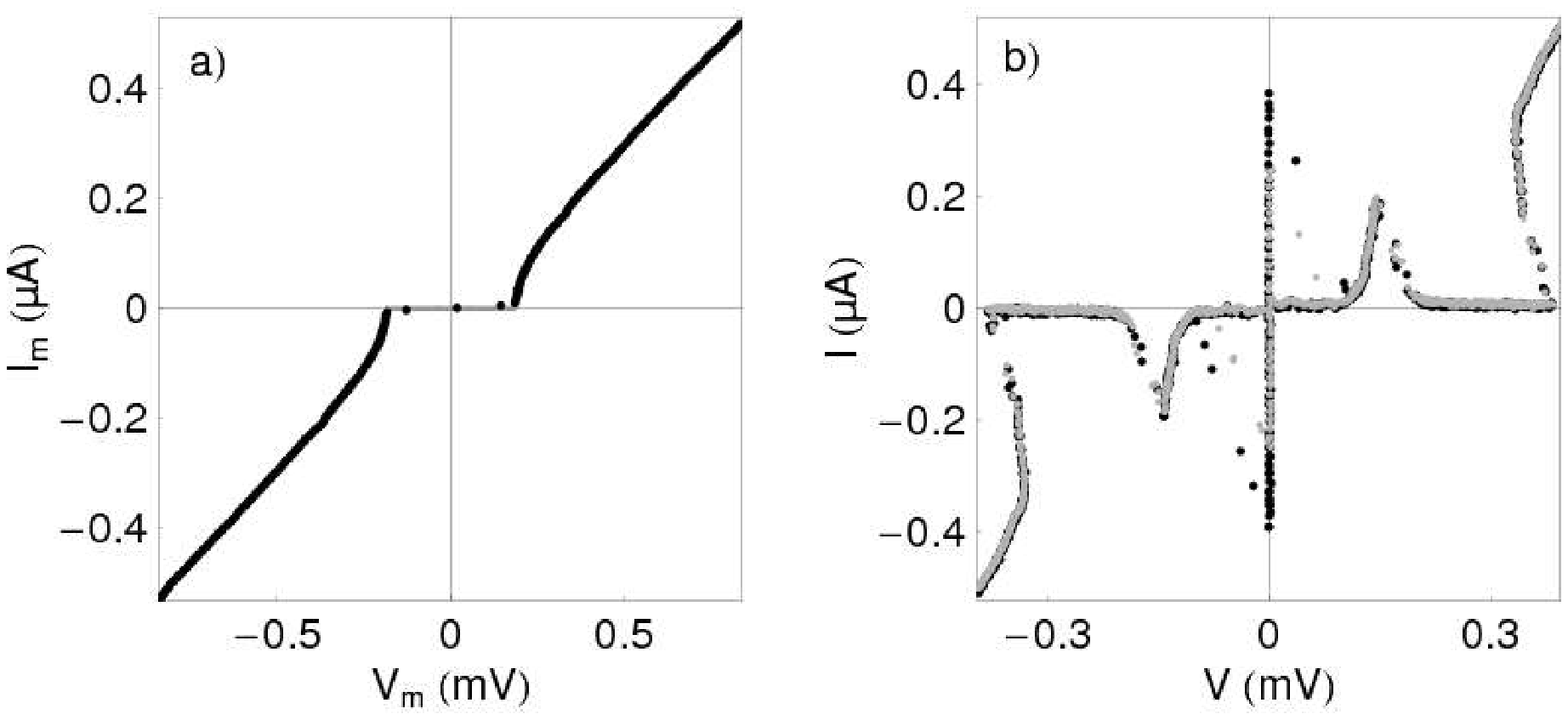}
\end{center}
\vchcaption{a) Current-voltage characteristics of the NIS junction
at a refrigerator temperature of $T_{0}=20~\mathrm{mK}$. A fit to
theory yields the tunnel resistance of
$R_{t}=1.57~\mathrm{k}\varOmega$ and the superconducting gap
$\varDelta=185.5~\protect\mu \mathrm{eV}$.b) Current-voltage
characteristics of the Josephson junction at two different values of
the current $I_{m}$ in the NIS junction and at a refrigerator
temperature of $T_0=20~\mathrm{mK}$. The black dots were measured
with $\langle I_{m}\rangle =5.3~\mathrm{nA}$ and the grey dots with
$\langle I_{m}\rangle =640~\mathrm{nA}$. } \label{IVJJ}
\end{vchfigure}

The current-voltage characteristics of the Josephson junction, shown
in Fig.\thinspace \ref{IVJJ} for two values of the current through
the NIS\ junction, is obtained by sweeping back and forth the
voltage $V_{b},$ and recording both the current (from the voltage
across a $10\,\Omega $ resistor placed at the base temperature) and
the voltage.\ A supercurrent branch is observed, and when
$V_{b}/R_{b}$ exceeds a threshold $I_{sw}$, the junction switches.\
The current then drops essentially to zero, with the applied voltage
$V_{b}=R_{b}I_{sw}\sim 0.1\,\mathrm{mV}<2\Delta /e$ now entirely
across the Josephson junction. Compared to a perfect current bias,
our bias scheme limits the production of quasiparticles at the
Josephson junction. However, we observed in the $I-V$
characteristics of the Josephson junction a large sub-gap current
peak close to $V\sim 0.15\,\mathrm{mV}$, leading to increased
heating for the largest switching current. We therefore had to
reduce the duty cycle of the applied pulses to a value such that
heating could be neglected. In order to extract the parameters
determining the dynamics of
the Josephson junction, we first used resonant activation \cite%
{resonant,theseBH} to extract the plasma frequency $\omega _{0}\simeq 1.0\,%
\mathrm{GHz}$ at $s=0$, and determine its quality factor
$Q_{0}\simeq 22$. In addition, from the value of the plasma
frequency, one can estimate the cross-over temperature between the
regimes of thermal activation and quantum
tunneling: $T_{\mathrm{co}}=\hbar \omega _{p}/2\pi k_{B}\approx 8~\mathrm{mK}%
\ll T_{0}$. Therefore, and as stated above, switching does occur due
to thermal activation for all experimentally accessible
temperatures. Then, from the measurement of the number of switching
events for a large number of current pulses, we deduced the
dependence of the switching rate on the current through the
Josephson junction. Equation (\ref{rate}) then yields the critical
current $I_{0}\simeq 0.48\,\mu \mathrm{A}$ and the escape
temperature $T_{\mathrm{esc}}\simeq 35\,\mathrm{mK}$. The fact that
the latter temperature is slightly larger than the fridge
temperature comes
mainly from the heating of the resistor $R_{b}$ by the current pulses \cite%
{theseBH}.

\begin{vchfigure}[tbph]
\begin{center}
\includegraphics[width=11.5cm]{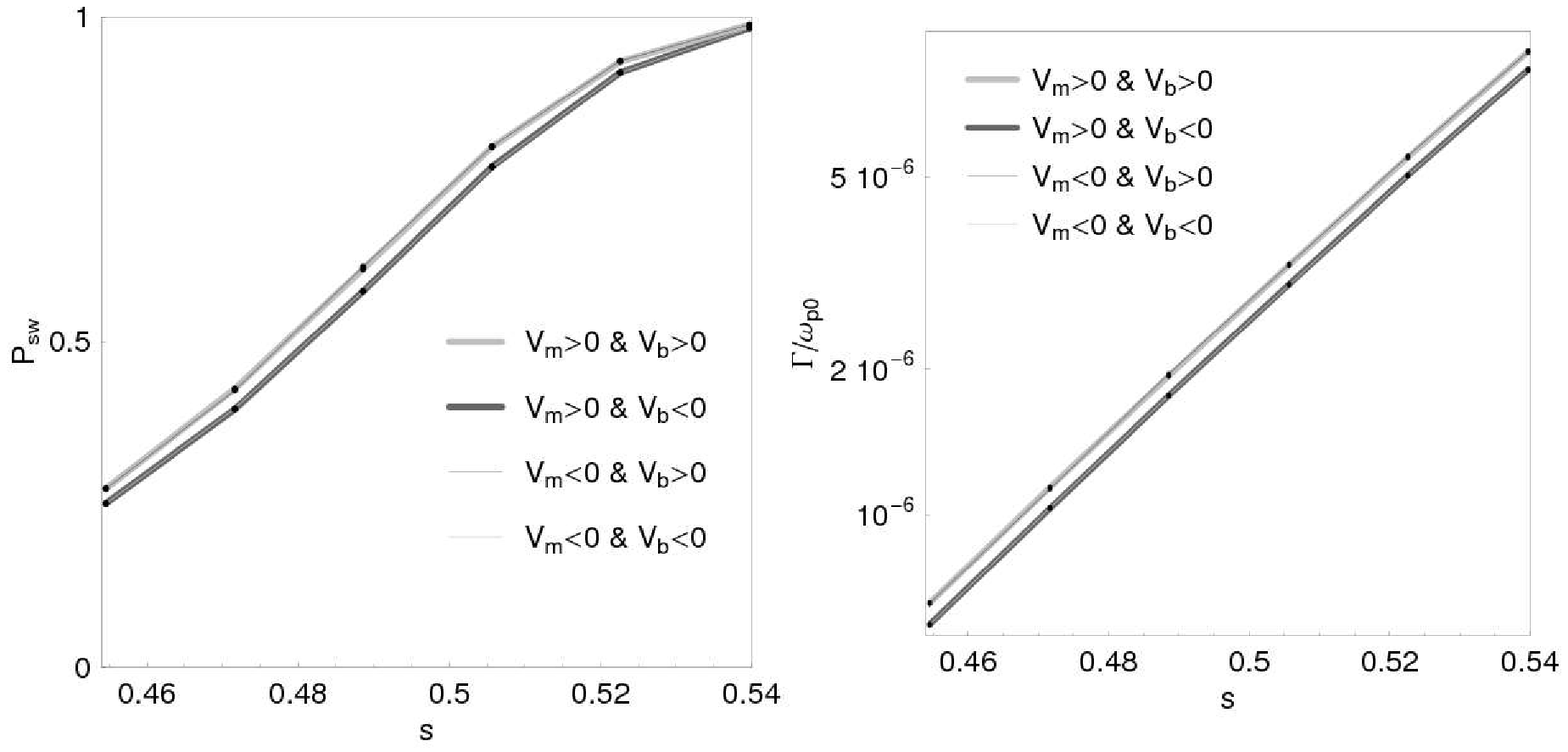}
\end{center}
\vchcaption{Switching probabilitiy $P_{\mathrm{sw}}$ (left pannel)
and corresponding switching rates $\Gamma $ (right pannel) as a
function of the
reduced pulse height $s=I_{b}/I_{0}$ for $\langle I_{m}\rangle \approx 1.1~%
\protect\mu \mathrm{A}$ at a temperature $T_{0}\approx
197~\mathrm{mK}$. The probabilities are calculated from statistics
on $2\times 10^{5}$ to $5\times 10^{5}$ pulses in order to achieve a
constant relative incertainty $5\times 10^{-8}\protect\omega _{0}$
on the rate $\Gamma $ (see Appendix), smaller than the size of the
dots. The four curves correspond to four combinations
of the signs of $V_{m}$ and $V_{b}.$ Pairs with identical sign of $%
V_{m}\times V_{b}$ are almost superimposed, as expected from
time-reversal symmetry.} \label{Scurve}
\end{vchfigure}

For non-zero values of the average current $\langle I_{m}\rangle $
through the NIS junction, where the statistics of tunneling events
are expected to be Poissonian, the switching properties of the
Josephson junction are probed as follows. Sequences of $10^{4}$
current pulses with alternating polarity and of duration $\tau
_{p}\approx 50~\mu \mathrm{s}$ are applied to the Josephson
junction. In order to circumvent artifacts associated with possible
DC offsets and thermoelectric effects, the current pulses are sent
through a large, room temperature capacitor ($200\,\mu \mathrm{F}$).
For each pulse, the voltage across the Josephson junction is
monitored by two threshold detectors, each detecting the switching
events in one of the two current directions (appearance of a
positive or a negative voltage larger than a fixed threshold).
Between each sequence of $10^{4}$ pulses, the polarity of the
current $\langle I_{m}\rangle $ through the NIS junction is
inverted. The switching probabilities $P_{sw}$ can then be
reconstructed for 4 combinations of the signs of $I_{m}$ and
$I_{b}.$ It was carefully checked that the combinations $(I_{m}>0,$
$I_{b}>0)$ and $(I_{m}<0,$ $I_{b}<0),$ on the one hand, and
$(I_{m}>0,$ $I_{b}<0)$ and $(I_{m}<0,$ $I_{b}>0),$ on the other
hand, give the same results, as expected from symmetry
considerations. As an example, the dependence of $P_{sw}$ on
$s=\left| I_{b}\right| /I_{0}$ is shown in Fig.\thinspace
\ref{Scurve} for $\left\langle I_{m}\right\rangle =\pm
1.1\,\mathrm{\mu A}.$

This measurement is then repeated for various pulse heights, and for various
$I_{m}.$ The range of pulse heights was chosen to keep the relative
uncertainty on the switching rates small (see appendix). Information on the
average noise and its asymmetry are extracted from these data. We first
focus on the average noise, which is the largest effect, seen in the escape
temperature $T_{\mathrm{esc}}$ according to Eq.~(\ref{rate}).

From $P_{sw}$, one extracts the switching rate $\Gamma =-\tau _{p}^{-1}\log
(1-P_{sw})$ and the curve $\Gamma (s)$ is fitted with Eq.\thinspace (\ref%
{rate}). The only fitting parameter is the escape temperature $T_{\mathrm{esc%
}}$, which does not change for opposite current directions within
the experimental error margins. The dependence of $T_{\mathrm{esc}}$
on $\left\langle I_{m}\right\rangle $
is shown in Fig.\thinspace \ref{Tesc}, for bath temperatures $T_{0}$ from $%
20\,\mathrm{mK}$ to $530\,\mathrm{mK.}$
\begin{vchfigure}[tbph]
\begin{center}
\includegraphics[width=7cm]{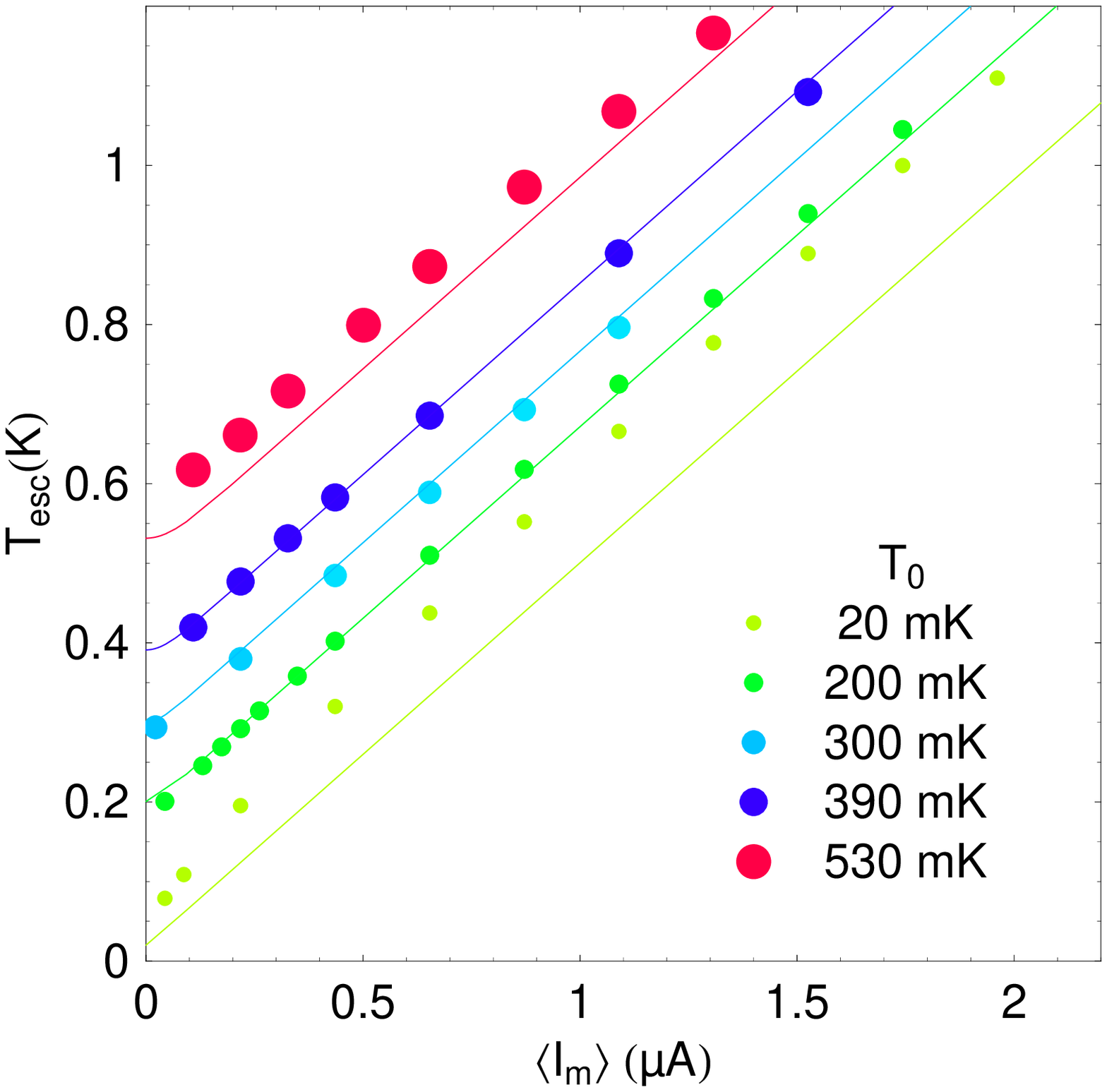}
\end{center}
\vchcaption{Escape temperatures extracted using Eq.~\ref{rate} (with $I_{0}=479~%
\mathrm{nA}$, $C=36~\mathrm{pF}$) from the measurement of the rate as a
function of $s$ for several values of the current in the tunnel junction $%
I_{m}$ and of the refrigerator temperature $T_{0}$. The lines show the
behavior expected from Eq.(\ref{Tescpredicted}), in which $a=eR_{\parallel
}/2k_{B}$ was taken as a fit parameter ($a_{\mathrm{fit}}\simeq 0.48\,%
\mathrm{K/\protect\mu A}$).}
\label{Tesc}
\end{vchfigure}

Apart from the $20\,\mathrm{mK}$ data, one finds (lines in Fig.\thinspace %
\ref{Tesc}) $T_{\mathrm{esc}}\simeq 0.95\,T_{0}+a\left\langle
I_{m}\right\rangle ,$ with $a\simeq 0.48\,\mathrm{K/\mu A}$. This
slope can be compared to the prediction (\ref{teff de noise}), in
which $R$ has
to be replaced with the parallel combination of $R_{f}$ and $R_{b}$: at $%
eR_{\parallel }\left\langle I_{m}\right\rangle \gg 2k_{B}T_{0},$ one expects
\begin{eqnarray}
T_{\mathrm{esc}} &=&R_{\parallel
}(1/R_{f}+1/R_{b})T_{0}+eR_{\parallel }\left\langle
I_{m}\right\rangle /2k_{B}~\mathrm{coth}\left[ eR_{t}\langle
I_{m}\rangle /2k_{B}T_{0}\right] \nonumber   \\
&\approx &0.95\,T_{0}+eR_{\parallel }\left\langle I_{m}\right\rangle
/2k_{B},\label{Tescpredicted}
\end{eqnarray}%
with $R_{\parallel }=(1/R_{f}+1/R_{b}+1/R_{t})^{-1}$, leading to
$a\simeq 0.55\,\mathrm{K/\mu A}$, in reasonable agreement with the
value observed. The slight difference in $a $ can be attributed to
an imperfect modeling of the impedance seen by the Josephson
junction near $\omega _{p}/2\pi \approx 1~\mathrm{GHz}$, where it is
the most sensitive. The increased escape temperature at
$T_{0}=20~\mathrm{mK}$ is well explained by electron heating in the
resistors $R_{f}$ and $R_{b}$ in presence of the currents $I_{m}$
and $I_{b}$ \cite{theseBH,heating}.

The switching of a Josephson junction is not only sensitive to the average
noise power of current fluctuations, but also to higher order cumulants. The
contribution of the cumulants with odd order yield an asymmetry of the
switching rate $R_{\Gamma }(s)=\Gamma _{+}(s)/\Gamma _{-}(s)-1,$ with $%
\Gamma _{+}(s)$ the rate when $I_{b}$ and $I_{m}$ have the same sign, $%
\Gamma _{-}(s)$ when they have opposite signs. As discussed in the
appendix, we acquired data only for a  $0.25<P_{sw}<0.987,$ an
interval in which a good precision on $R_{\Gamma }$ can be reached
with achievable numbers $N$ of measurement pulses. Since
$T_{\mathrm{esc}}$
increases with $\left\langle I_{m}\right\rangle ,$ data taken at various $%
\left\langle I_{m}\right\rangle $ correspond to different ranges of
$s.$ For example, at 20~mK, $s\approx 0.9$ at $\langle
I_m\rangle=0.04~\mu\mathrm{A}$ and $s\approx 0.3$ at $\langle
I_m\rangle=2~\mu\mathrm{A}$. This limits the accessible range for
the current $I_m$.

The existence of an asymmetry can be seen already in the raw data (Fig.~\ref%
{Scurve}). Figure \ref{rgammaFIG} shows, for various temperatures $%
T_{0},$ the dependence of the asymmetry $R_{\Gamma }$ on
$\left\langle I_{m}\right\rangle $ for the two extreme values of
rate exponents $B=9.5$ and $B=11.5$, corresponding to $P_{sw}\approx
0.25$ and $P_{sw}\approx 0.987,$ and to switching rates $\Gamma
\approx 6~\mathrm{kHz}$ and $\Gamma\approx 90~\mathrm{kHz}$. In each
case, the two curves are very similar, with $R_{\Gamma}$ increasing
continuously with $\left\langle I_{m}\right\rangle \mathrm{.}$

\begin{vchfigure}[tbph]
\begin{center}
\includegraphics[width=12.5cm]{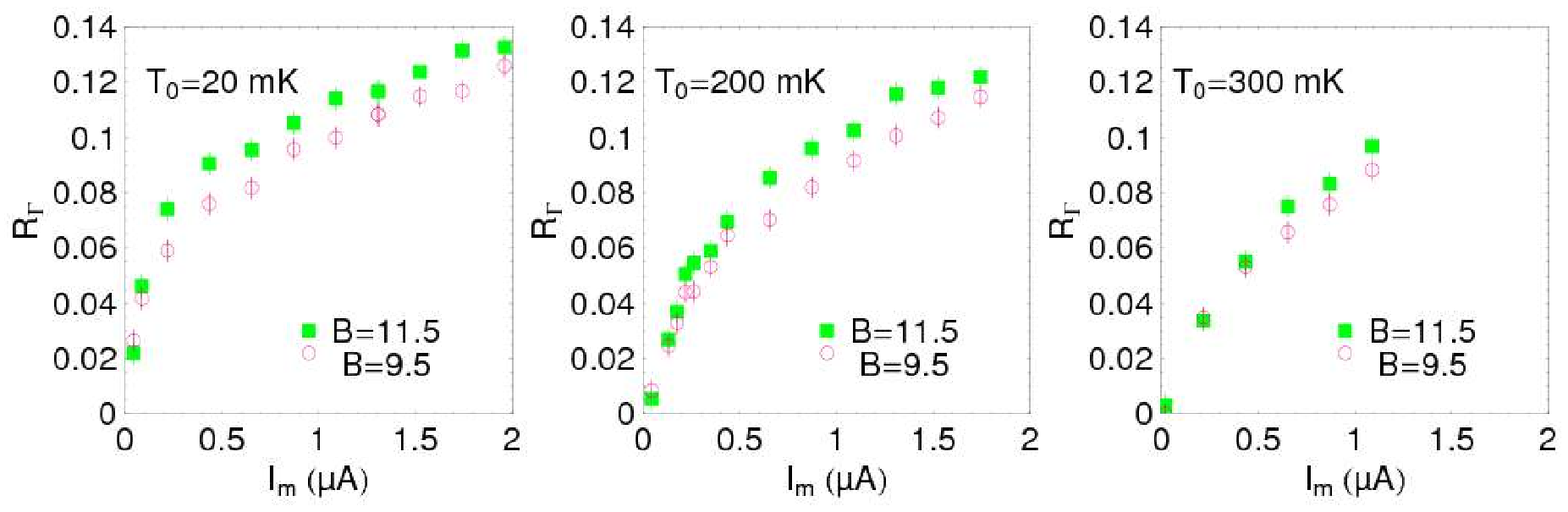}
\end{center}
\vchcaption{Measured asymmetry $R_\Gamma$ between positive and
negative bias schemes as a function of the mean current $I_m$
through the tunnel junction. Each panel corresponds to a single
refrigerator temperature $T_0$ and each symbol corresponds to a
fixed value of the rate exponent $B$ from Eq.~(\ref{rate}). Error
bars are shown as lines (see Appendix).} \label{rgammaFIG}
\end{vchfigure}

In order to compare these results with theory, we first discuss how the
switching of a Josephson junction is affected by non-gaussian white noise.

\section{Dynamics of a Josephson junction in presence of non-Gaussian noise}

In this section we present a theoretical analysis of the switching
of a Josephson junction out of its zero voltage state driven by
thermal as well as weak non-Gaussian noise. According to the
experimental situation described above (Fig.~\ref{SchemeEPS.eps}),
the underlying stochastic processes are assumed to be classical,
both for the dynamics of the detector, the Josephson junction, and
for the electrical noise originating from the mesoscopic conductor,
the tunnel junction. Hence, the standard RSJ model applies, where
now the total noise consists of the Johnson-Nyquist noise $\delta
I$, in the simplest case produced by a resistor $R$ in parallel to
the Josephson junction, and of the mesoscopic noise $\delta I_{m}$
from the fluctuating part of the mesoscopic current $I_{m}=\langle
I_{m}\rangle +\delta I_{m}$. Since the dc-component $\langle
I_{m}\rangle $ of the mesoscopic current can be
incorporated trivially, we only focus in the sequel on its fluctuating part $%
\delta I_{m}$. Then, according to the Kirchhoff rules we have for the
currents entering the detector
\begin{equation}
I_{b}+\delta I_{m}(t)=\frac{V}{R}+I_{0}\sin \varphi +C_{J}\dot{V}\,,
\end{equation}%
where $\varphi $ denotes its phase difference and the other quantities are
defined according to Fig.~\ref{SchemeEPS.eps}. To proceed it is convenient
to work with dimensionless quantities and to scale times with $\omega _{0}$,
energies with $E_{J}$, and currents with $I_{0}$. This way, we arrive with $%
\dot{\varphi}=V/\varphi _{0}$ and $\tau =\omega _{0}t,$ at a classical
Langevin equation for the phase dynamics, i.e.,
\begin{equation}
\ddot{\varphi}=-\sin \varphi -\frac{\dot{\varphi}}{Q_{0}}+s_{b}+s_{\mathrm{m}%
}(\tau )+s_{\mathrm{th}}(\tau )  \label{eq:reduced}
\end{equation}%
with the quality factor $Q_{0}=RC_{J}\omega _{0}$ of the unbiased junction,
the scaled bias current $s_{b}=I_{b}/I_{0}$, and the scaled noise sources $%
s_{\mathrm{th}}=\delta I/I_{0}$ and $s_{\mathrm{m}}=\delta
I_{m}/I_{0}$ for thermal and non-Gaussian noise, respectively. These
two noisy forces are assumed to be Markovian, which is an accurate
approximation since the experiment is operated in a regime where the
typical correlation times are much smaller than the typical time
scales of the junction, e.g.\ the inverse of the plasma frequency.
This situation also ensures that we can work with
continuous noise processes. Hence, we have for the thermal noise $\langle s_{%
\mathrm{th}}\rangle =0$ and
\begin{equation}
\langle s_{\mathrm{th}}(\tau )s_{\mathrm{th}}(0)\rangle =\frac{2\theta }{%
Q_{0}}\,\delta (\tau )\,,  \label{thermsecond}
\end{equation}%
where we introduced the dimensional temperature $\theta =k_{\mathrm{B}%
}T_{0}/E_{J}$. The non-Gaussian noise produced by the tunnel junction is
taken as Poissonian with $\langle s_{\mathrm{m}}\rangle =0$, where
specifically the second moment reads in the high voltage/low temperature
limit $(eR_{t}I_{m}\gg 2k_{B}T_{0})$
\begin{equation}
\langle s_{\mathrm{m}}(\tau )s_{\mathrm{m}}(0)\rangle =F_2\alpha ^{2}\Lambda _{%
\mathrm{m}}\delta (\tau )\,,  \label{mesosecond}
\end{equation}%
with $\Lambda _{\mathrm{m}}=\langle I_{m}\rangle /I_{0}\alpha $, $\alpha
=e\omega _{0}/I_{0}$ and the third moment is given by
\begin{equation}
\langle s_{\mathrm{m}}(\tau _{2})s_{\mathrm{m}}(\tau _{1})s_{\mathrm{m}%
}(0)\rangle =F_3\alpha ^{3}\Lambda _{\mathrm{m}}\delta (\tau
_{2})\delta (\tau _{1})\,.  \label{thirdmoment}
\end{equation}%
The prefactors $F_2$ and $F_3$ are the Fano factors for the second
and third cumulant. Here, for the tunnel junction, $F_2=F_3=1$. It
is important to realize that the average mesoscopic current $\Lambda
_{\mathrm{m}}$ can be large so that the second moment
(\ref{mesosecond}) can
be comparable to or larger than that of the thermal noise (\ref{thermsecond}%
), while the third moment is still small due to $\alpha \ll 1$. In fact, as
discussed above, the Gaussian component of the mesoscopic noise effectively
adds to the thermal noise to determine the effective temperature of the
junction
\begin{eqnarray}
\theta _{\mathrm{eff}} &=&\theta +\theta _{\mathrm{m}}  \notag \\
&\equiv &\theta +F_2\frac{\alpha ^{2}\Lambda
_{\mathrm{m}}Q_{0}}{2}\,. \label{efftemp}
\end{eqnarray}

So far we have assumed that the resistance seen by the Josephson
junction is exclusively given by the resistance $R$ in parallel to
it so that (\ref{efftemp}) agrees with (\ref{teff de noise}) only in
the limit $R_t\gg R$. In a description closer to the experimental
situation the total resistance is $R R_t/(R+R_t)$ and we have to
replace $\theta_{\mathrm{eff}} \to R_t/(R+R_t)\,
\theta_{\mathrm{eff}}$, thus arriving precisely at (\ref{teff de
noise}). Another simplification is that in the Langevin equation
(\ref{eq:reduced}) no back-action effect from the detector to the
tunnel junction has been taken into account. The back-action appears
because when the Josephson junction switches, the voltage across the
tunnel junction no longer coincides with the external voltage, but
is reduced by $(\hbar/2e) \langle\dot{\varphi}\rangle$. Accordingly,
on the one hand in the friction term in (\ref{eq:reduced}) $Q_0 $
has to be replaced by $R_t/(R+R_t) Q_0$ and on the other hand the
third moment (\ref{thirdmoment}) carries for finite $e
V/\theta_{\mathrm{eff}}$ an
additional contribution from the derivative of the second moment (\ref%
{mesosecond}) with respect to the voltage. For a detailed quantitative
comparison with the experiment these modifications must be taken into
account.

Based on (\ref{eq:reduced}) we first give a brief account of a theoretical
approach developed recently by one of us \cite{JA}, where an analytical
expression for the rate and its asymmetry due to a weak third cumulant of a
current through a mesoscopic conductor has been derived. In a second part,
direct numerical simulations are presented, which allow for a detailed
comparison with the analytical predictions and may thus lay out a firm basis
for the comparison between theory and experiment.

\subsection{Analytical rate expression}

As usual in rate theory, for analytical treatments it is much more
convenient to work with phase space probabilities rather than
individual stochastic trajectories. The general problem for
non-Gaussian noise is then that the corresponding Fokker-Planck
equation (FPE) based on a Kramers-Moyal expansion contains diffusion
coefficients up to infinite order. The basic idea for weak
non-Gaussian noise with a leading third cumulant is thus to derive
an effective, finite order Fokker-Planck equation. Based on a
cumulant expansion of the noise generating functional such a
generalized FPE has been derived in \cite{JA} and leads to a FPE
with a momentum dependent diffusion term, where the momentum
dependence is weighted by the third cumulant.

The existence of a time-independent switching rate is associated
with a quasi-stationary nonequilibrium distribution, which reduces
to a thermal equilibrium in the well region and deviates from it
only around the barrier top in the case where friction is moderate
to high. This in turn means that the
barrier height compared to the thermal energy must be sufficiently large $%
\Delta U/k_{\mathrm{B}} T_{\mathrm{eff}}\gg 1$ so that switching becomes a
rare event on the time scale of the plasma frequency. The typical thermal
rate expression then looks as in (\ref{rate}) and is dominated by the
exponential (activation factor) being identical to the probability to reach
the barrier top from the well bottom by a thermal fluctuation. Accordingly,
the quasi-stationary state (the flux state) is written as a product
consisting of the thermal distribution and a 'form factor' describing the
deviation from it around the barrier top. The latter one leads to the rate
prefactor, while the former one specifies the activation factor. Here, we
restrict ourselves to the exponential factor, which provides by far the
dominant contribution in the present experimental set-up. Within this
theoretical framework in \cite{JA} an analytical expression for the exponent
including leading corrections due to a third cumulant have been obtained.
The rate takes the form $\Gamma_\pm\propto \exp[-B (1\mp |g|)]$ with the
dimensionless barrier height $B=\Delta U/(E_J \theta_{\mathrm{eff}})$ and
the correction $g$ due to the third cumulant such that $\Gamma_+$
corresponds to $\Lambda_{\mathrm{m}}>0$ and $\Gamma_-$ to $\Lambda_{\mathrm{m%
}}<0 $. Note that we thus consider the bias current $s_b$ as being always
positive, while the situation in which one keeps $\Lambda_{\mathrm{m}}>0$
and inverts the bias current follows straightforwardly.

In the dimensionless quantities introduced above one has
\begin{equation}
B=\frac{4\sqrt{2}}{3\theta _{\mathrm{eff}}}(1-s_{b})^{3/2}\,.  \label{Bthy}
\end{equation}%
The rate asymmetry $R_{\Gamma }=\Gamma _{+}/\Gamma _{-}-1$ is found as \cite%
{remark}
\begin{equation}
R_{\Gamma }\approx \exp \left[ F_3 \frac{16\sqrt{2}\alpha ^{3}\,|\Lambda _{%
\mathrm{m}}|}{9\,\theta _{\mathrm{eff}}^{3}}\frac{Q^{2}}{5+Q^{2}}\frac{%
(1-s_{b})^{5/2}}{\sqrt{1-s_{b}^{2}}}\right] -1  \label{rateasymm}
\end{equation}%
with the quality factor of the biased junction
$Q=Q_{0}(1-s_{b}^{2})^{1/4}$. For weak friction $Q\gg 1$, the only
dependence of $R_{\Gamma }$ on
friction is through the effective temperature (see Eq.\thinspace (\ref%
{efftemp})). When the asymmetry is small enough, it is directly
proportional to the Fano factor $F_3$, leading to a clear measure of
the third order cumulant of the noise (Fig.~\ref{fig:numerical}b).

Let us first look at the dependence of $R_\Gamma$ when the
mesoscopic current varies (Fig.~\ref{fig:numerical}a). For small
$\Lambda_{\mathrm{m}}$ the effective temperature coincides with the
actual temperature $\theta_{\mathrm{eff}}\approx \theta$
and $R_\Gamma$ grows linearly with $\Lambda_{\mathrm{m}}$. For large $%
\Lambda_{\mathrm{m}}$ the heating originates solely from the mesoscopic
current, i.e.\ $\theta_{\mathrm{eff}}\approx \theta_{\mathrm{m}}\propto
\Lambda_{\mathrm{m}}$, so that $R_\Gamma$ decreases roughly $\propto
1/\Lambda_{\mathrm{m}}^2$ and the rate asymmetry vanishes due to a growing
second cumulant compared to the third one. Consequently, there is a
changeover between these two regimes with a maximum of the rate asymmetry
around the range, where $\theta\approx\theta_{\mathrm{m}}= \alpha^2 \Lambda_{%
\mathrm{m}} Q_0/2$ [see (\ref{efftemp})].

In the experiment, the junction is operated essentially at a fixed
switching rate, with $s_{b}$ tuned at each mesoscopic current
$\Lambda _{\mathrm{m}}$ to a value such that $B$ is kept constant.
In this situation it is instructive to look for the scaling
properties of the rate asymmetry as a function of the bare exponent
$B$. In general, to express $R_{\Gamma }(s_{b}) $ in terms of $B$
leads to a rather involved expression. Transparent formulae are
found if the asymmetry is sufficiently small so that a linearization
of the exponential in (\ref{rateasymm}) applies, for vanishing
thermal noise $\theta =0$, and in the limits of strong and weak
friction, respectively. Then, one writes for (\ref{rateasymm})
$R_{\Gamma }(B)\approx \xi _{\eta }\,B^{\eta }$ and looks for that
exponent $\eta $ for which the prefactor $\xi _{\eta }$ displays the
weakest dependence on $B$. This way, one finds in the regime of weak
friction that $\eta \approx 1.45$, while for strong friction one has
$\eta \approx 5/3$.

To complete this discussion, we remark that an analytical expression for the
asymmetry of the escape rate in the limits of low and high friction has been
also derived in \cite{Eugene} within the framework of an Onsager-Machlup
type of functional representation of the stochastic process combined with an
asymptotic evaluation (high barrier, weak third cumulant) of the path
integrals. The expression (\ref{rateasymm}) reduces for low $Q$ exactly to
the one specified there. For high $Q$ both results give also precisely the
same dependence on the physical quantities, but with slightly different
numerical constants. Further, back-action effects according to the
discussion at the end of the previous section can be included in (\ref%
{rateasymm}) in the same way as in Ref.~\cite{Eugene} since both
theoretical approaches start from the same Langevin equation. Hence,
we conclude that, up to small corrections, both theories make
identical predictions, where the advantage of (\ref{rateasymm}) is
that it provides an analytical expression for all damping strengths.

\subsection{Numerical simulations}

Direct numerical simulations of the escape rate of a Josephson junction in
presence of a weakly non-Gaussian noise are based on the Langevin equation (%
\ref{eq:reduced}) with the mesoscopic noise term $s_{\mathrm{m}}(t)$ being
represented as
\begin{equation}
s_{\mathrm{m}}(\tau)=\sum_i \alpha \delta(\tau-\tau_i) -\alpha\Lambda_{\mathrm{m}%
}\, .
\end{equation}
The random variables $\tau_i$ are the times at which a tunneling event
occurs in the mesoscopic tunneling junction. Those tunneling events
correspond to a Poisson process so that the intervals between two
consecutive events $S_i=\tau_{i+1}-\tau_i$ are independently distributed
according to $p(S_i)=\Lambda_{\mathrm{m}} e^{-\Lambda_{\mathrm{m}} S_i}$.

The numerical integration of Eq.(\ref{eq:reduced}) should be performed with
some care as the non-Gaussian effects we are looking for are rather weak. We
have therefore used two independent schemes and checked for several cases
that they produced identical results within statistical errors.

\begin{vchfigure}[tbp]
\begin{center}
\includegraphics[width=5.5cm]{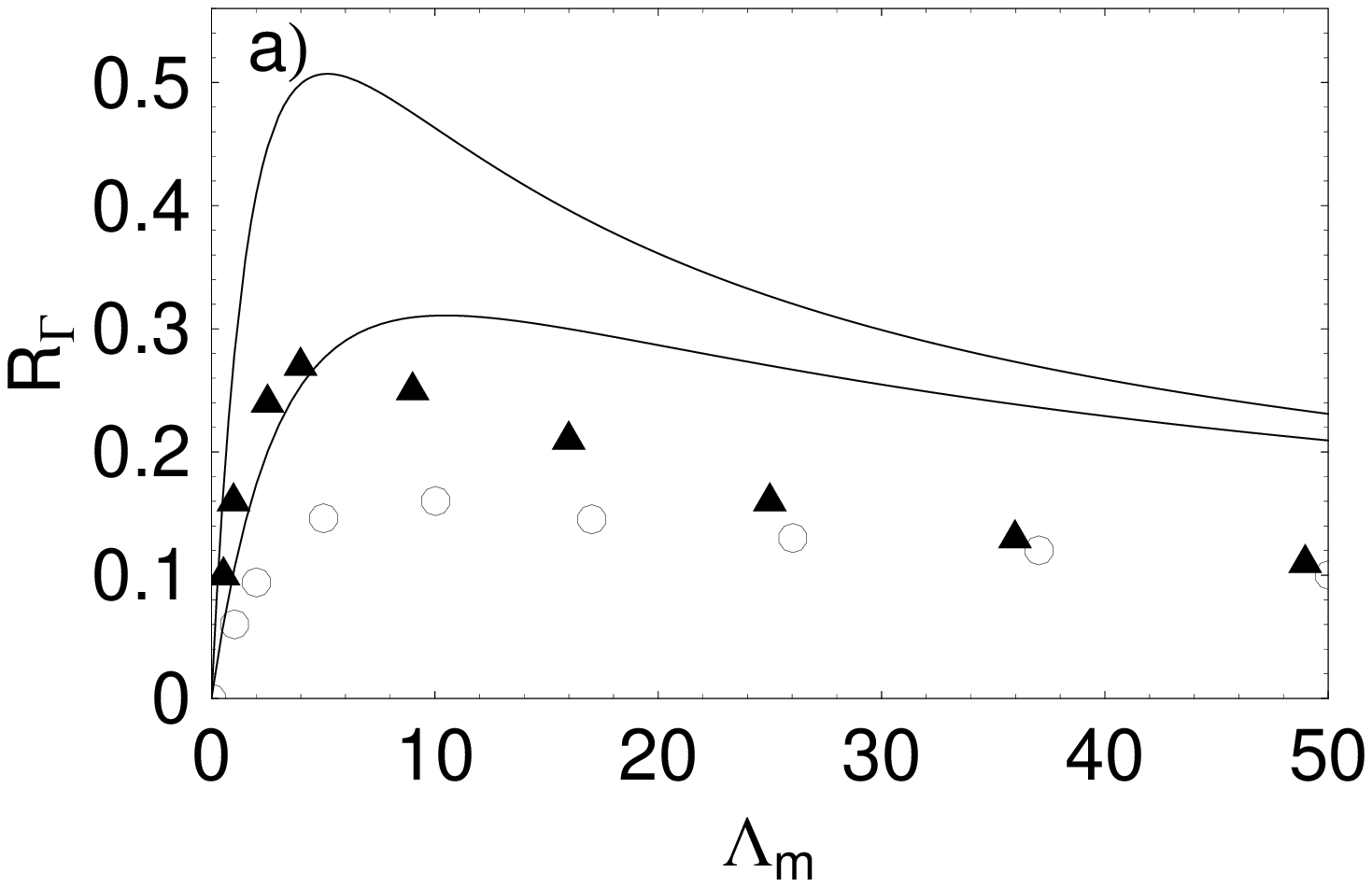} %
\includegraphics[width=5.5cm]{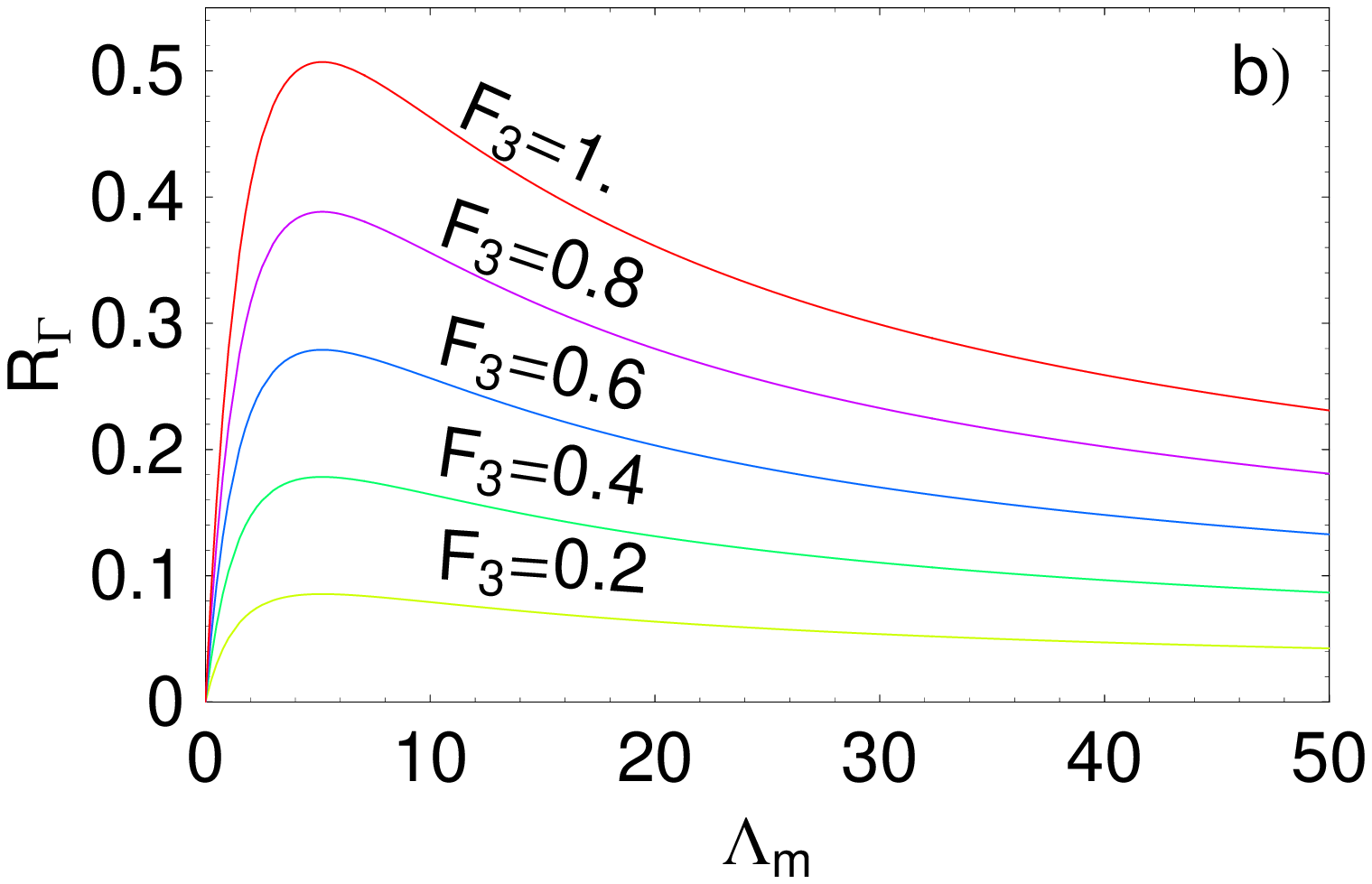} %
\includegraphics[width=5.5cm]{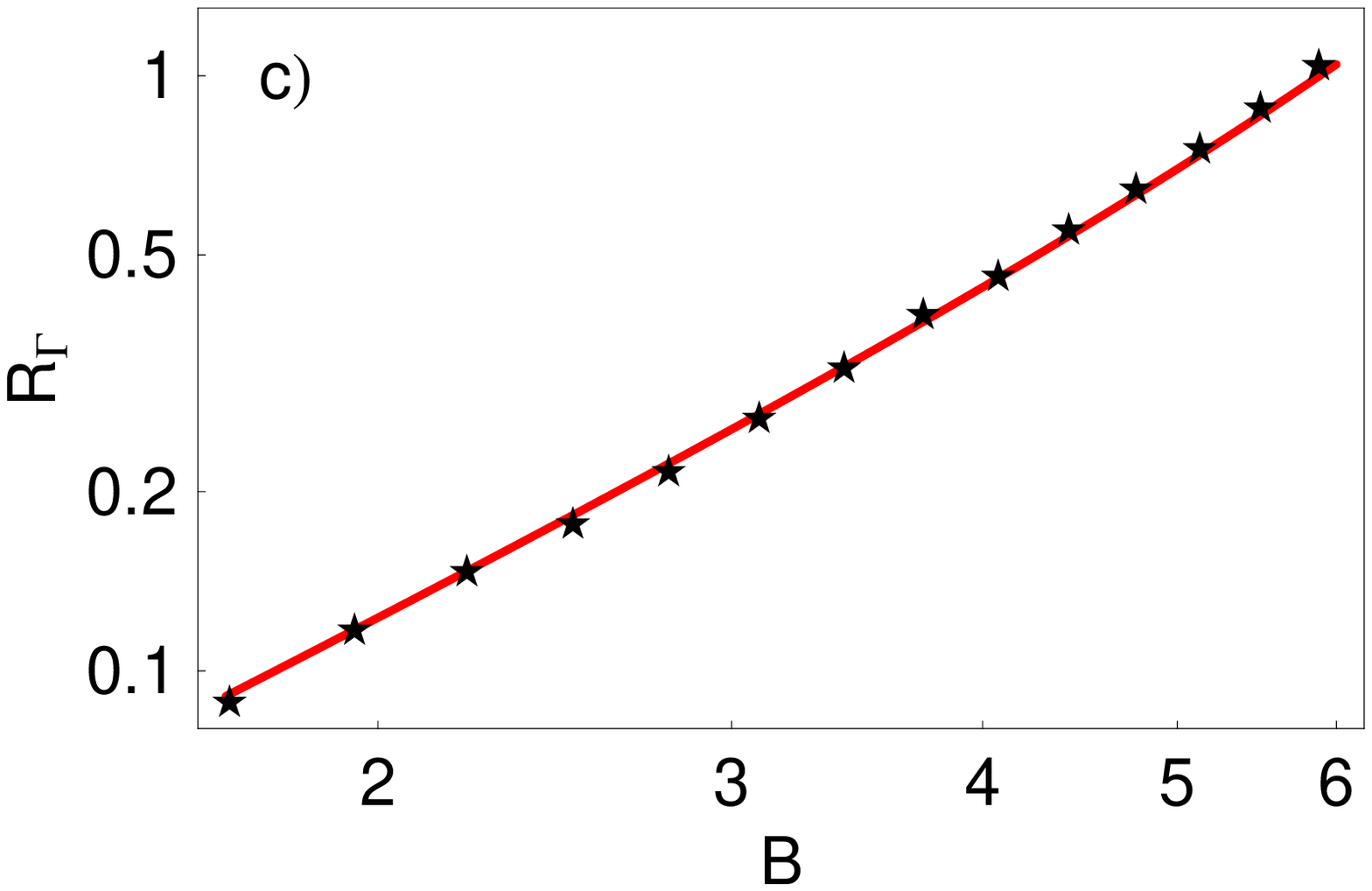} %
\includegraphics[width=5.5cm]{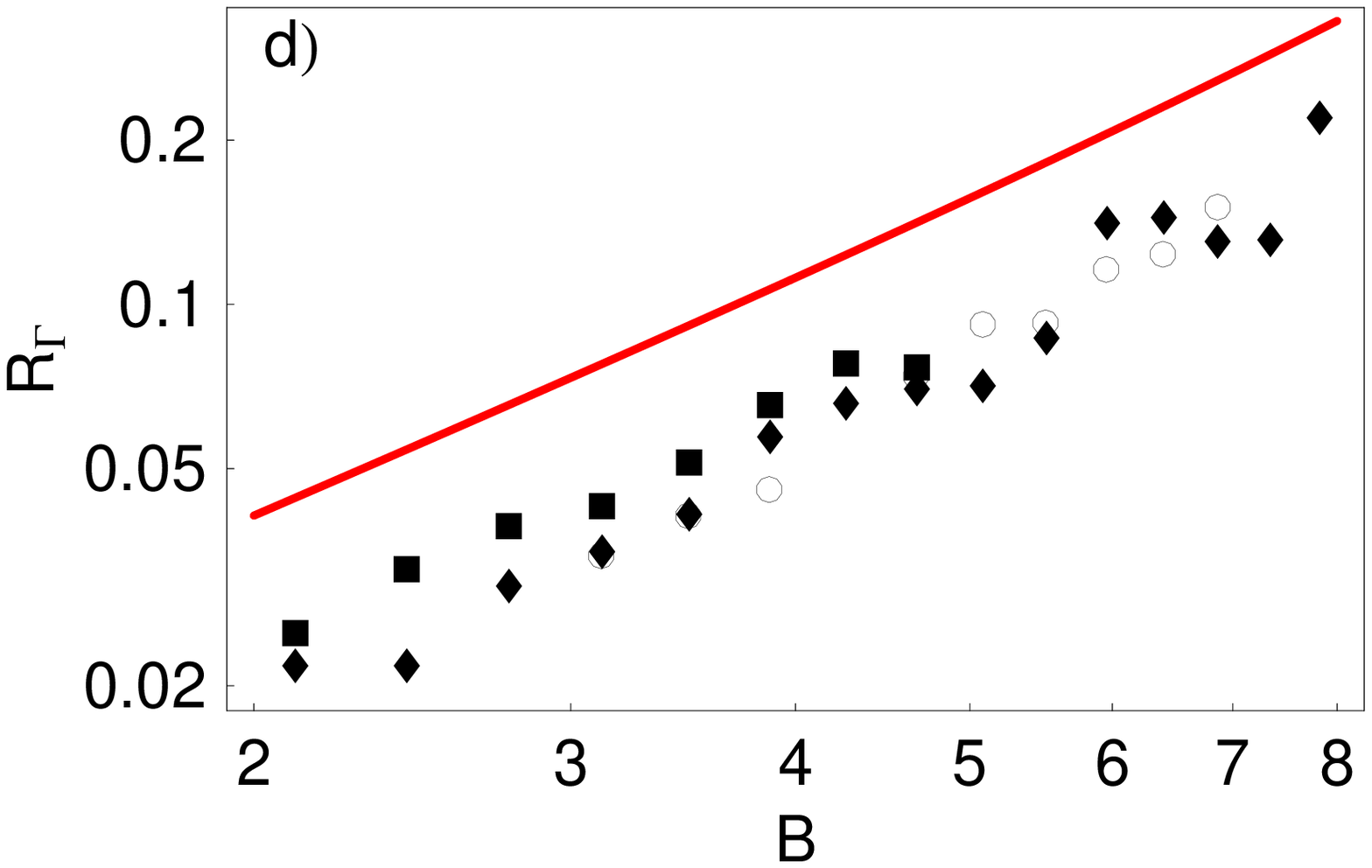}
\end{center}
\vchcaption{Numerical results versus theory. \textbf{a)} numerical
simulations for the asymmetry of the tunneling rates as a function
of the current $\Lambda _{\mathrm{m}}$ at a fixed value of the
exponent $B=6$ in the rate (\ref{rate}) and for two
values of the temperature $\protect\theta =0.005$ ($\blacktriangle $) and $%
\protect\theta =0.01$ ($\circ $). $Q_{0}=8$, $\protect\alpha =0.02$. The
maximum occurs at the crossover $\protect\theta _{\mathrm{m}}=\protect\theta $%
. The lines are plotted using Eq.~(\ref{rateasymm}) with the same
parameters. \textbf{b)} The lines are plotted using
Eq.~(\ref{rateasymm}) with the same parameters as in a) and
$\protect\theta =0.005$. Each line corresponds to a different Fano
factor for the third order cumulant of the noise. The experiment and
all the other figures describe the case $F_3=1$. \textbf{c)}
asymmetry as a function of the exponent $B$ for $%
Q_{0}=0.5$, $\protect\alpha =0.1$, $\protect\theta =0$ and $\Lambda _{%
\mathrm{m}}=40$. The line is plotted using Eq.~(\ref{rateasymm})
with the same parameters. \textbf{d)} asymmetry as a function of the
exponent $B$ for
$Q_{0}=10$, $\protect\alpha =0.02$, $\protect\theta =0$ and $\Lambda _{%
\mathrm{m}}=40$. The various symbols stand for the first integration scheme (%
$\circ $: $N=10^{5}$, $h=0.006$) and second integration scheme ($%
\blacksquare $: $N=10^{6}$, $h=0.08$ ;$\blacklozenge $: $N=10^{5}$,
$h=0.0044 $). The line is plotted using Eq.~(\ref{rateasymm}) with
the same parameters. } \label{fig:numerical}
\end{vchfigure}

\subsubsection{First scheme}

The first scheme is a first order direct integration of the equation of
motion. We discretize the time into steps of size $h$. Noting $%
\varphi_i=\varphi(\tau=i h)$ and $p_i=\dot\varphi(\tau=i h)$, we
have
\begin{eqnarray}
p_{i+1}-p_i &=& (F_i -p_i/Q_0) h+\Sigma_i^{\mathrm{m}} +\Sigma_i^{\mathrm{th}%
}  \notag \\
\varphi_{i+1}-\varphi_i &=& p_i h
\end{eqnarray}
with
\begin{equation*}
F_i=-\sin \varphi_i + s \ \ \ \mathrm{and } \ \ \
\Sigma_i^\Upsilon=\int_{i h}^{(i+1)h} d\tau \ \ s_\Upsilon(\tau) \ \
\ (\Upsilon=\mathrm{m ; th})
\end{equation*}
The variables $\Sigma_i^{\mathrm{th}}$ are independent and Gaussian of
variance $2\theta/(Q_0h)$. In this first scheme, we have directly sampled
the variables $S_i$ through $S_i=-\log(X_i/\Lambda_{\mathrm{m}})$ where $X_i$
is a random number uniformly distributed within $[0,1]$. $\Sigma_i^{\mathrm{m%
}}= \alpha \hat N -\alpha\Lambda_{\mathrm{m}}$ counts the number $\hat N$ of
tunneling events in the window $[i h,(i+1)h]$.

\subsubsection{Second scheme}

The second scheme is a direct extension of Ref.~\cite{linkwitz1992} and is
valid up to second order in $h$,
\begin{eqnarray}
p_{i+1}-p_i &=& (F_i+F_{i+1})h/2 +\Sigma_i^{\mathrm{m}}+\Sigma_i^{\mathrm{th}%
} -(\varphi_{i+1}-\varphi_i)/Q_0  \notag \\
\varphi_{i+1}-\varphi_i &=& p_i h \left[1-h/(2 Q_0)\right] +F_i h^2/2
+(\Sigma_i^{\mathrm{m}}+\Sigma_i^{\mathrm{th}}) h/2\, .
\end{eqnarray}
There, we have directly sampled the number $\hat N$ of tunneling events
which is distributed according to $P(\hat N= i)=e^{-\Lambda_{\mathrm{m}}
h}\, \frac{(\Lambda_{\mathrm{m}} h)^i}{i!}$.

\subsubsection{Numerical results}

To compute the escape rate $\Gamma$, we integrate the equations of motions $%
N $ times (typically $N=10^5$) and save the time $\tau_j$ where
$\varphi$ escapes its well (taken as $|\varphi|>3.5$). We checked
that the $\tau_j$ are
distributed according to an exponential law $P(\tau_j<\tau)=1-e^{-\Gamma \tau}$%
. Noting $\tau_p$ the maximal time of one simulation, and $<n>
\approx e^{-\Gamma \tau_p}$ the measured probability of non
switching, our estimator for $\Gamma$ is $\frac{1}{\Gamma}=
\frac{<\tau_j>}{1-<n>}$. The asymmetry of
the escape rate is defined as $R_\Gamma=\Gamma(\Lambda_{\mathrm{m}%
})/\Gamma(-\Lambda_{\mathrm{m}})-1$.

For the discussion of our numerical findings let us start with a comparison
with the analytical predictions presented in the previous section. In Fig.%
\ref{fig:numerical}a the rate asymmetry as a function of the average
mesoscopic current $\Lambda _{\mathrm{m}}$ is depicted for fixed barrier
height $B=6$. In accordance with theory $R_{\Gamma }$ shows a non-monotonous
behavior with a maximum around that value of $\Lambda _{\mathrm{m}}$ where $%
\theta \approx \theta _{\mathrm{m}}$. For higher thermal noise the rate
asymmetry shrinks for lower $\Lambda _{\mathrm{m}}$, but becomes independent
of thermal noise in the decreasing tail for higher $\Lambda _{\mathrm{m}}$.
Apparently, while the analytical expression provides the overall tendency
correctly, there are quantitative deviations from the simulated data. The
reason for that will be discussed below.

The scaling of the rate asymmetry as a function of $B$ is shown in Fig.\ref%
{fig:numerical}c for strong ($Q_0=0.5$) and in
Fig.\ref{fig:numerical}d for weaker ($Q_0=10$) friction. In the
former case we find an accurate agreement with the analytical
expression (\ref{rateasymm}) even for low barriers. In the latter
situation the analytical expression describes the scaling with
respect to $B$ correctly, however, for lower barriers the absolute
values exceed the numerical ones by about 100\%, while for somewhat
higher barriers, for which convergent data could be obtained for
$6<B\leq 8$, the deviations become smaller, namely 50\%. Note, that
for weaker damping the variance of the numerical data around an
average behavior increases and is typically of the order of 10\%.
For even larger values of $B$ the switching becomes an extremely
rare event and the statistics deteriorate drastically. The obvious
discrepancy for lower barriers that has already been seen in
Fig.\ref{fig:numerical}a does not come as a surprise since, as known
from standard Kramers' rate theory, escape rates are calculated as
expansions in $1/B$ and the exponential activation factor dominates
only for sufficiently large $B$. In particular in the weak
friction regime finite barrier corrections as well as the rate \emph{%
prefactor} tend to play an even more important role when one approaches the
energy-diffusion limit, where the rate vanishes with decreasing friction
strength. The turnover between energy-diffusion and spatial-diffusion occurs
around $Q_0 \approx B$ so that the simulated data displayed in Fig.\ref%
{fig:numerical}d in the range of lower barrier heights are certainly
influenced by these effects.

The overall picture emerging is hence that within the statistical
uncertainties of the simulations, the analytical theories developed
for the switching of the Josephson junction predict quantitatively
the influence of a weak third cumulant of non-Gaussian noise up to
minor deviations due to prefactor effects over a broad range of
parameters. As expected, larger discrepancies occur for weaker
friction and lower barriers, a range beyond the strict applicability
of the analytical treatments.

\section{Discussion}

\begin{vchfigure}[hptb]
\begin{center}
\includegraphics[width=6.5cm]{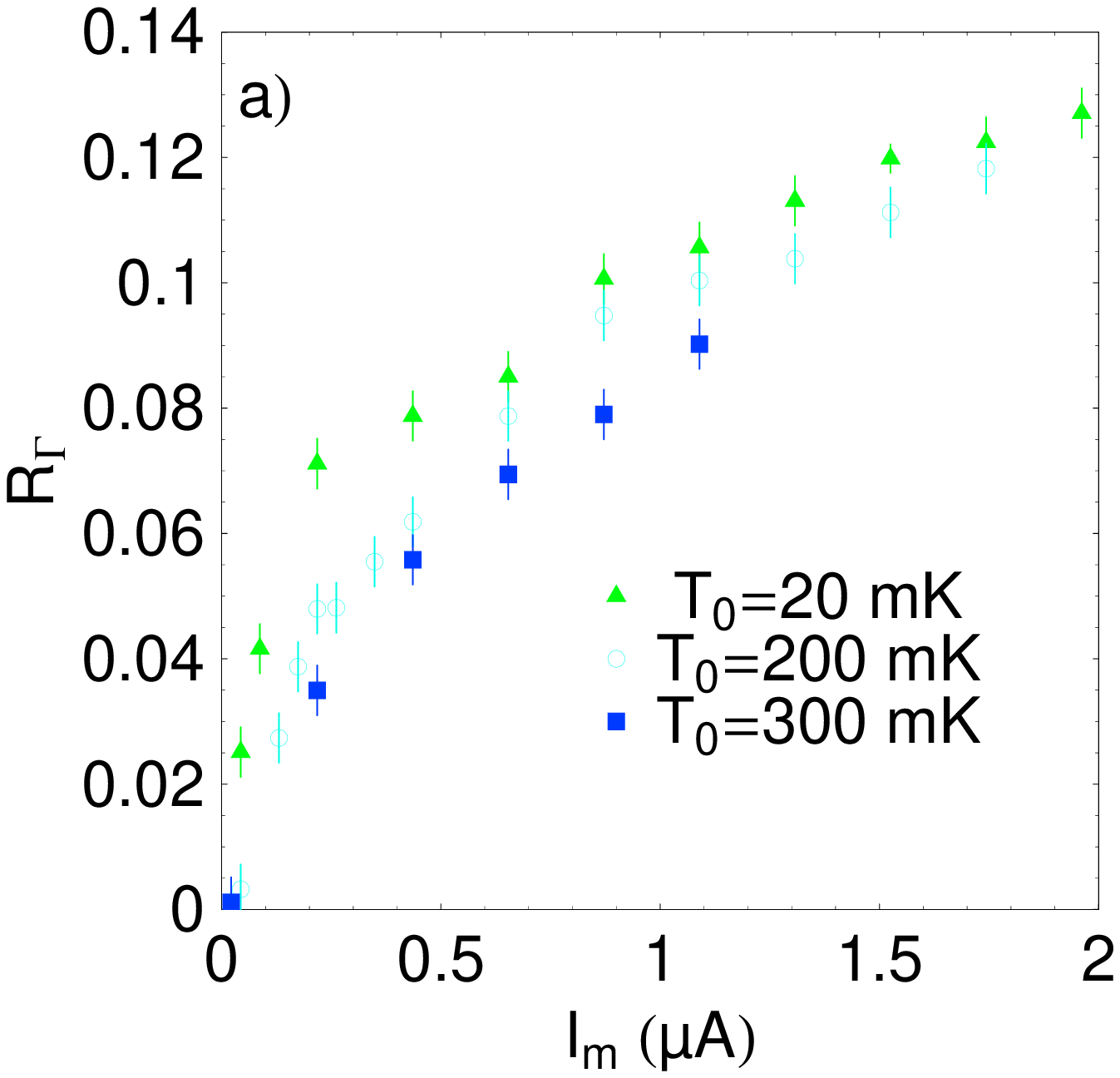}
\includegraphics[width=6.5cm]{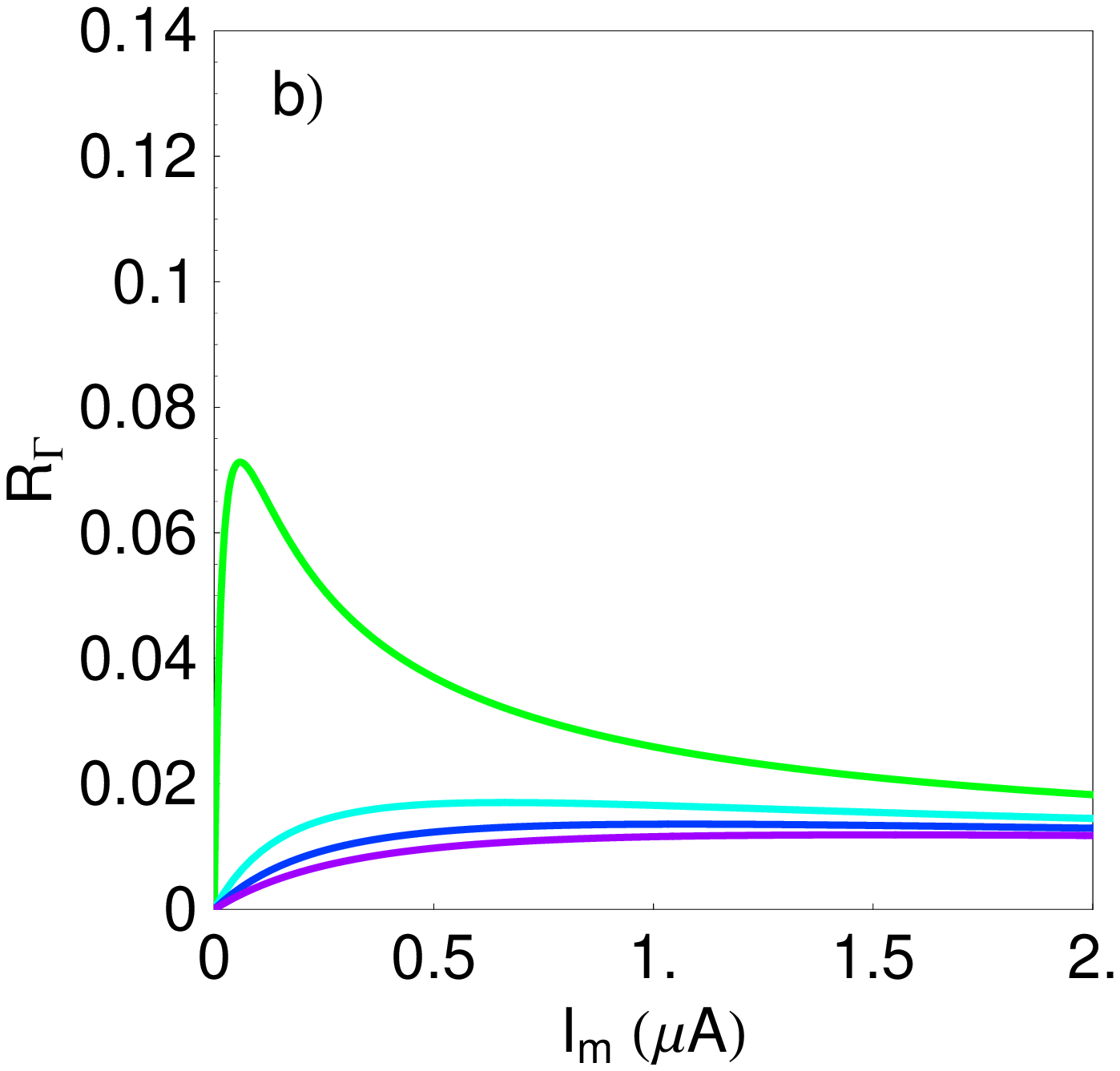}
\end{center}
\vchcaption{\textbf{a)} Measured asymmetry $R_\Gamma$ as function of
the
current $I_m$ at a fixed value of the exponent $B=10.5$ in the rate (\ref%
{rate}) for various temperature. \textbf{b)} Same quantity calculated using (%
\ref{rateasymm}), assuming the actual temperature is the refrigerator
temperature. }
\label{rgammascalingFIG}
\end{vchfigure}

Figure~\ref{rgammascalingFIG} presents a comparison between the measured and the predicted asymmetry in the switching
rates of the Josephson junction in presence of a current in the tunnel junction. The data of the left pannel (a)
correspond to the intermediate the switching exponent $%
B=10.5$. The right panel (b) shows the theoretical predictions Eq.~(\ref{rateasymm}) for the same parameters, and
assuming that the temperature is $T_{0}$. The disagreement is significant, both in the magnitude (asymmetries found in
the experiment are larger than in theory) and in the current-dependence. In particular, the increase of the asymmetry
with $\left\langle I_{m}\right\rangle $ is in strong contradiction with theory, which predicts a maximum when the shot
noise due to the tunnel junction is similar in magnitude to the thermal noise seen at zero current. Two reasons can be
invoked for this disagreement: an imperfect modeling of the experiment by theory, or experimental artifacts.

\subsection{Effects not considered by theory}
The experimental setup presents differences from the simplified circuit described by theory: the $RC$ combination that
allows in the experiment to separate low
and high frequency fluctuations is not treated by theory; the resistance $%
R_{f}$ in series with the tunnel junction also produces noise that modifies
the voltage across the tunnel junction, a configuration in which feed-back
effects lead to an asymmetry of the current fluctuations \cite{ReuletHouches}
that has not been considered; the voltage across the Josephson junction has
fluctuations due to plasma oscillations, leading to voltage fluctuations
that have been neglected. At least this last point can be discarded, since
the corresponding voltages $\approx \varphi _{0}\omega _{p}$ are in the $\mu
V$ range. More work in necessary to analyze the effects of the two other
main differences between experiment and theoretical modeling.

\begin{vchfigure}[htbp]
\begin{center}
\includegraphics[width=6.5cm]{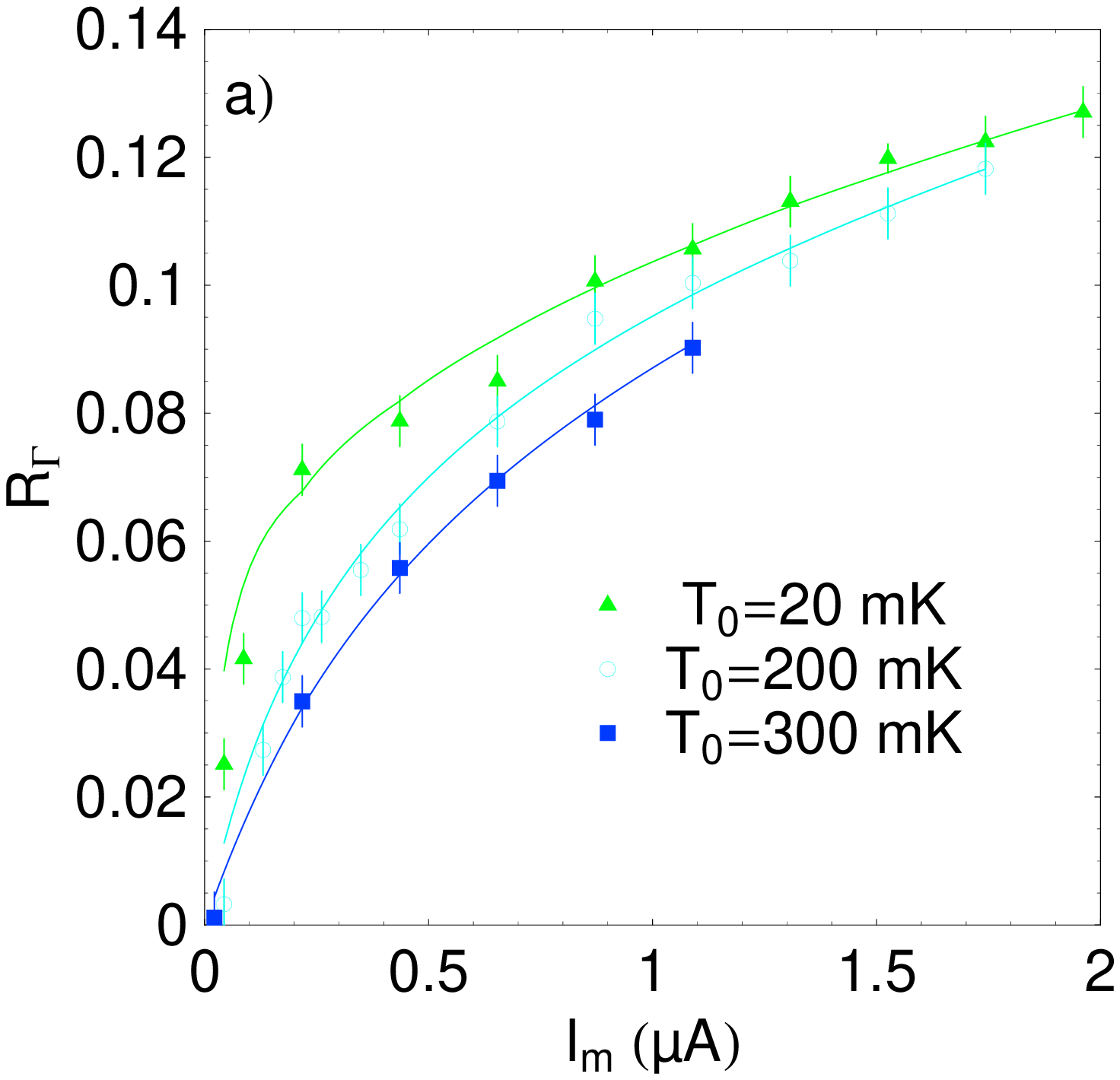} %
\includegraphics[width=6.5cm]{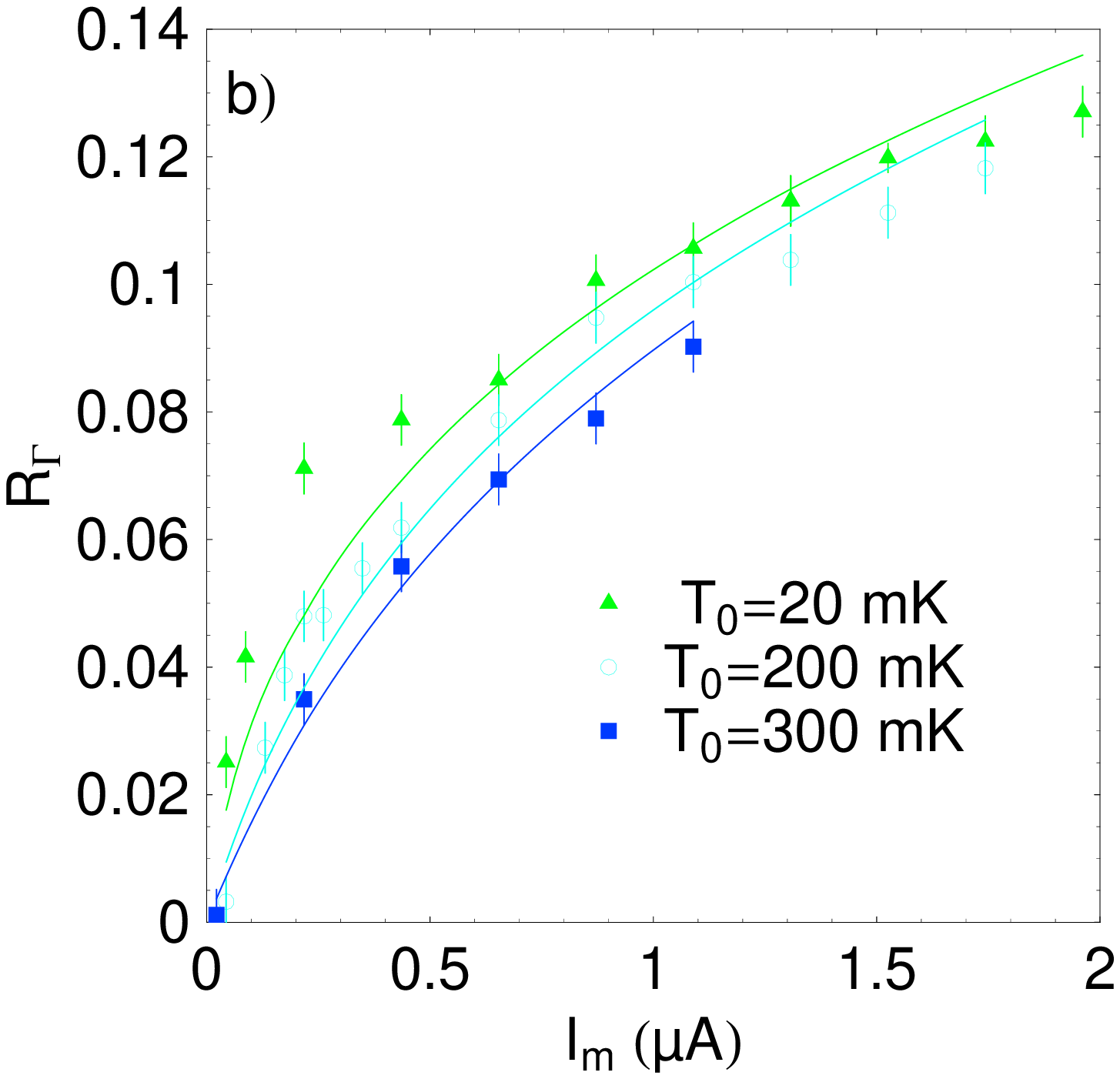}
\end{center}
\vchcaption{\textbf{a)} Plot of the asymmetry $R_\Gamma$ for a fixed value $%
B=10.5$ assuming that the actual value of the current $s$ is given by $%
s\leftarrow s\pm\protect\lambda \langle I_m\rangle/I_0$ because of a
possible leak through the capacitor $C_1$. Besides, the temperature $\protect%
\theta_\mathrm{eff}$ was extracted from Fig.\ref{Tesc} instead of
being calculated from Eq.(\ref{efftemp}). Here,
$\protect\lambda=5.5\times 10^{-4}$ leads to the best fit with the
data. \textbf{b)} Same thing assuming that the third moment is 0,
that is $g=0$. Here, $\protect\lambda=7\times 10^{-4}$ leads to the
best fit with the data.} \label{totoleretour}
\end{vchfigure}

\subsection{Experimental artifact?}
We have also explored another possible source of asymmetry associated with an hypothetic imperfection in the
experiment, namely a leak across one of the coupling capacitances $C$ or $C_{1}$ or a touch to ground of the
corresponding bias line. In this case, a fraction $\lambda $ of the DC current through the junction passes through the
Josephson junction instead of returning entirely through the resistor $R_{f}$: at $\left\langle I_{m}\right\rangle >0$,
the switching rate at $I_{b}>0$ is increased, whereas at at $\left\langle I_{m}\right\rangle <0$ it is decreased. The
corresponding asymmetry can be evaluated using Eq.~(\ref{rate}) with $B$ evaluated at $s=(I_{b}\pm \lambda \left\langle
I_{m}\right\rangle )/I_{0}.$ For $\lambda \ll 1$, one obtains
\begin{equation}
R_{\Gamma }\approx 3B(1-s)^{-1}\lambda \left\langle
I_{m}\right\rangle /I_{0}.
\end{equation}%
The best fit of the data with this expression in which the measured escape temperatures are plugged in, obtained with
$\lambda =7\times 10^{-4}$, is presented in Fig.~\ref{totoleretour}b. The general trend is well reproduced, but
systematic deviations are observed for the data at $T_{0}=20~\mathrm{mK}$ and at currents above $1~\mu A$. In
Fig.~\ref{totoleretour}a, we fitted the data with the conjugated effect of a leak and of the asymmetry expected from
the third order cumulant of the current fluctuations (replacing again the effective temperature $\theta
_{\mathrm{eff}}$ by the measured escape temperature). In the best fit, obtained with $\lambda =5.5\times 10^{-4},$ all
the data are precisely accounted for. If this scenario is the correct one, the asymmetry due to the tunnel junction is
clearly seen in the data at the lowest bath temperature,
but the uncertainty is too large for other temperatures. The value of $%
\lambda $ corresponds to a leak resistance of $R_{t}/\lambda \approx 3~%
\mathrm{M}\Omega $ ($R_{f}/\lambda \approx 0.4~\mathrm{M}\Omega $) if the leak comes from the capacitor $C_{1}$ ($C$).
The capacitors were tested after fabrication, but a leak might have appeared afterwards. A touch to ground of the
corresponding connection line at this intermediate value seems unlikely.

\section{Conclusions} In conclusion, we have measured the switching rate of a Josephson junction when coupled to a
voltage-biased NIS junction. The main effect is a well explained increase of the escape temperature due to shot noise
through the junction. The asymmetry of the rates for opposite signs of the bias of the NIS junction is larger than
expected from the dynamics of the Josephson junction in presence of non-Gaussian noise. Part of the discrepancy might
be due to an oversimplification in the modelling of the circuit. It is however suspected that a current leak through a
capacitor, is responsible for a large fraction of the signal. Experiments less prone to experimental artifacts, with an
improved decoupling of the detector from the noise source at DC, are in progress.

\appendix

\section{Appendix: Sensitivity of the switching probability to a rate asymmetry}

For each current pulse we recorded the boolean $1$ or $0$ depending
whether the voltage across the Josephson junction crossed the
threshold or not. The booleans are $N$ independent random variables $%
X_{i}$ with binomial distribution, $P_{sw}$ being the expectation for $%
X_{i}=1$. Therefore, the standard deviation of the average $%
N^{-1}\sum_{i=1}^{N}X_{i}$ is $\sqrt{P_{sw}(1-P_{sw})/N}$. And using
\begin{equation}
P_{sw}=1-\mathrm{e}^{-\Gamma \tau _{\mathrm{p}}},
\end{equation}%
one gets the standard deviation of the switching rate
\begin{equation}
\delta \Gamma =\sqrt{\frac{P_{sw}}{N(1-P_{sw})}}\tau _{\mathrm{p}}^{-1}.
\label{errorrate}
\end{equation}%
Therefore, the uncertainty on the measurement of $R_{\Gamma }$ is given by
(see Eq.~(\ref{errorrate})):
\begin{equation}
\delta R_{\Gamma }=2\frac{\delta \Gamma }{\Gamma }=-2\sqrt{\frac{P_{sw}}{%
N(1-P_{sw})}}\frac{1}{\ln (1-P_{sw})}.  \label{accuracy
rgamma}
\end{equation}%
This function presents a minimum at $\delta R_{\Gamma }\approx 2.5/\sqrt{N}$
for $P_{sw}\approx 0.8$, and grows rapidly near $P_{sw}=0$ and $P_{sw}=1$.
In the experiment, we chose the range of $s$ so that $0.25<P_{sw}<0.987$ (or
equivalently $9.3<B<12$ in the notations of Eq.~(\ref{rate}))for which $%
\delta R_{\Gamma }<4/\sqrt{N}$, and adapted $N$ at each value of $P_{sw}$ to
keep $\delta R_{\Gamma }$ at a fixed value.

\begin{acknowledgement}
Discussions with Hermann Grabert, Jukka Pekola, Eugene Sukhorukov
and within the Quantronics group are greatfully acknowledged. This
work was partly funded by the Agence Nationale de la Recherche under
contract ANR-05-NANO-039. NB acknowledges partial support by the NSF
under grant DMR-0405238. JA acknowledges support from the DFG
through the Heisenberg program.
\end{acknowledgement}


\begin{thebibliography}{99}
\bibitem{remark} We note that in \cite{JA} a factor 2/3 has been forgotten
in the definition of the third cumulant. The result for the rate asymmetry
given there must thus be multiplied by 2/3 to get (\ref{rateasymm}).
\end{thebibliography}
\end{document}